# Experimental studies of pedestrian flows under different boundary conditions

Jun Zhang and Armin Seyfried

***Abstract*** **In this article the dynamics of pedestrian streams in four different scenarios are compared empirically to investigate the influence of boundary conditions on it. The Voronoi method, which allows high resolution and small fluctuations of measured density in time and space, is used to analyze the experiments. It is found that pedestrian movement in systems with different boundary conditions (open, periodic boundary conditions and outflow restrained) presents various characteristics especially when the density is larger than 2 m$^{-2}$. In open corridor systems the specific flow increases continuously with increasing density till 4 m$^{-2}$. The specific flow keeps constant in systems with restrained outflow, whereas it decreases from 1 (m.s)$^{-1}$ to zero in system with closed periodical condition.**

## I. INTRODUCTION

In the last few decades, several investigations have been done on pedestrian and traffic flow [1-5]. The study on pedestrian movement could support the safety of pedestrians in complex buildings or at mass events. The density-flow relationship, the so-called fundamental diagram, is one of the most important characteristics in pedestrian dynamics and is usually used in facility design and assessment. Nearly all methods in guidelines and handbooks like SFPE [6], Predtechenskii and Milinskii [7], Weidmann [8] are based on the assumption that there is a unique density-flow relation for corridors with or without bottlenecks and other narrowings. Several researchers have collected empirical data on the fundamental diagram [7-10]. Surprisingly, the small number of available datasets shows considerable disagreement [11, 12]. Reference [13] shows that the values for the even maximal density (3.8 to 10 m$^{-2}$) as well as capacity (1.75 to 7 m$^{-2}$) are different in various literatures.

Facing such problems, series of well-controlled laboratory experiments [14-19] as well as field studies [20-22] have been carried out in recent years. Based on these empirical results, several assumptions including cultural factors [23] and differences between unidirectional and multidirectional flow [24, 25] have been documented explaining some of the discrepancies mentioned above. In this study, we analyze laboratory experiments in open and closed corridors, T-junctions and bottlenecks to investigate the effect of boundary conditions. The structure of the paper is as follows. In Section 2 we describe the setup of the experiments. The analysis methodology and main results will be exhibited in Section 3. Finally, the conclusions will be discussed.

Jun Zhang, Jülich Supercomputing Centre, Forschungszentrum Jülich GmbH, Jülich 52428, Germany (corresponding author to provide phone: +49 24616196554; fax: +49 2461616656; e-mail: ju.zhang@fz-juelich.de).

Armin Seyfried, Jülich Supercomputing Centre, Forschungszentrum Jülich GmbH, Jülich 52428, Germany(e-mail: a.seyfried@ fz-juelich.de).

## II. EXPERIMENTAL SETUP

In this section, pedestrian experiments in four scenarios (straight and rounded corridor, T-junction and bottleneck) will be introduced. The first three experiments were carried out with up to 400 participants in hall 2 of the fairground Düsseldorf (Germany) in 2009, while the bottleneck experiment was performed with a group of soldiers in the wardroom of the Bergische Kaserne Düsseldorf in 2006. The whole processes of the experiments were recorded by cameras and the pedestrian trajectories are extracted automatically using *PeTrack* [26].

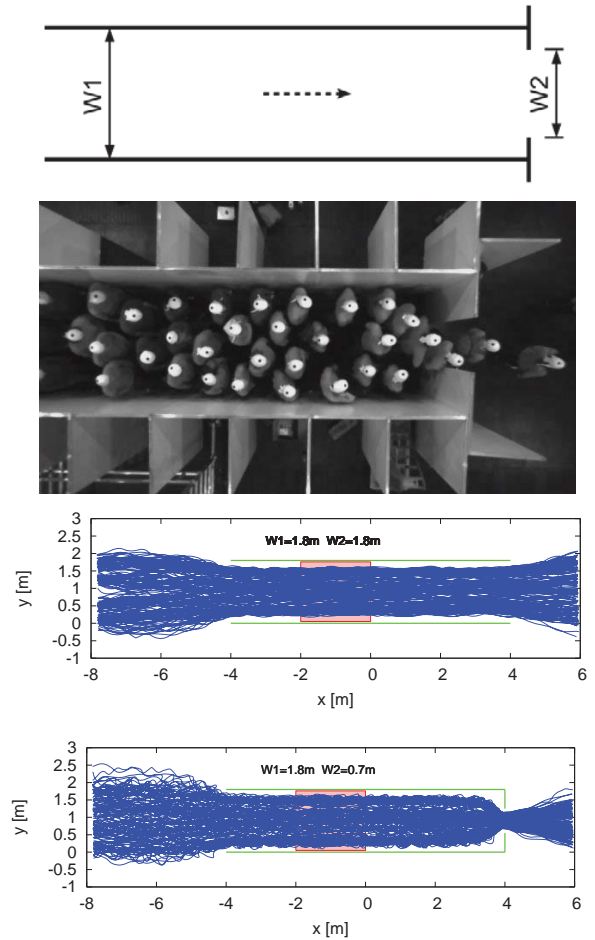

Figure 1. Sketch, snapshpot and pedestrian trajectories of the experiment in straight corridors. The squares in the on the trajectories represent themeasurement areas for calculating the density and velocity.

## A. Straight Corridor

Figure 1 shows the sketch and a snapshot of pedestrian movement in a straight corridor. Three corridor widths (W1=1.8 m, 2.4 m and 3.0 m) were selected to study the specific flow concept. To obtain high densities in corridor, the width of the exit (W2) was changed. Details of this experiment are presented in [27]. When W1 = W2, the movement at the end of the corridor is free and the outflow depends on the inflow. Yet when W1 > W2, the outflow is restrained and mainly depends on the width of exit. The densities in the corridor could be increased by decreasing the exit width.

## B. Rounded Corridor with Closed Boundary

Figure 2 shows the sketch, snapshot and trajectories of the experiment in a rounded corridor which has a closed and periodical boundary condition. Several runs with different corridor widths (W=1.0 m, 1.4 m and 1.8 m) and different numbers of participants in the corridor were performed. The participants were asked to move several rounds at a normal speed without hurry. Due to the special experimental conditions, only half of the scenario was recorded by cameras, which is seen from the pedestrian trajectories in the figure.

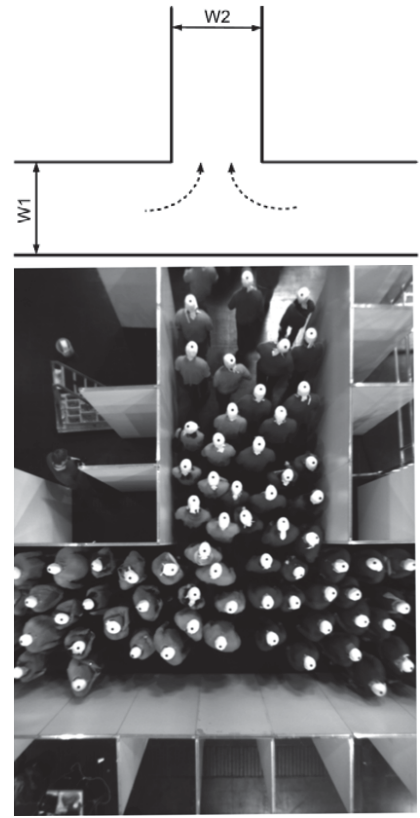

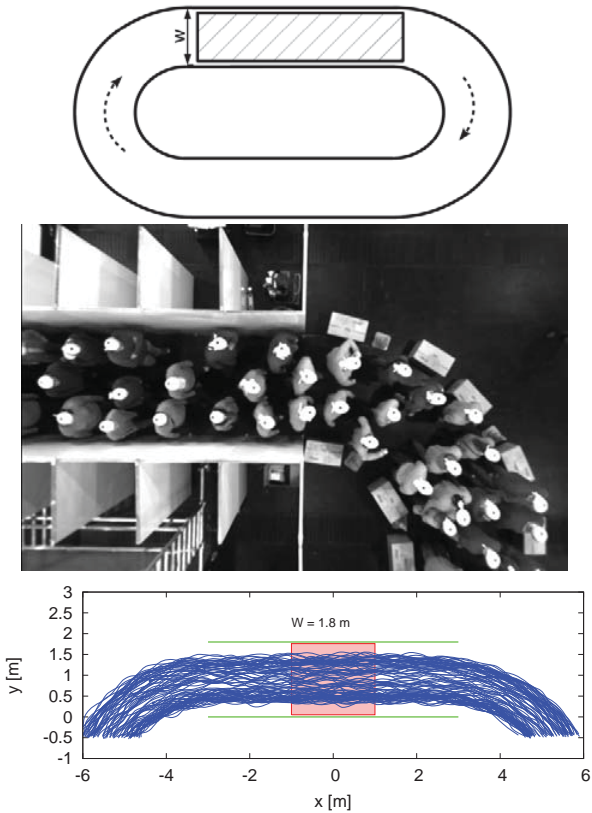

Figure 2. Sketch, snahspot and pedestrian trajectories of the experiment in rounded corridors with closed boundary condition.

Figure 3. Sketch, snahspot and pedestrian trajectories of the T-junction experiment.

## C. T-junction

The third scenario is merging flow in T-junctions with corridor widths W1 = W2 = 2.4 m and 3.0 m respectively. Figure 3 shows the sketch and trajectories from one run of the experiment. Two pedestrian streams move towards each other and join in the junction and form a single stream. The inflow rates of the two branches are approximately equal and regulated by changing the width of entrances each run. The number of participants in each run is set to a value so that the overall duration of all experiments was similar and was long enough to assure a steady state. The details for this experiment are presented in [27]. In this scenario, the outflow from the main stream is free. However, pedestrian movement in the junction is restrained by the turning and merging behavior as well as the reduction of the effective corridor width (from 2×W1 to W2 = W1).

*D. Bottleneck*

The last scenario is a bottleneck. Figure 4 shows the sketch, snapshot and trajectories from the bottleneck experiment. For a fixed bottleneck length (4 m), the bottleneck width W2 was changed from 0.9 m to 2.5 m. Pedestrian began to move from a waiting area to ensure an equal initial density (2.6 m$^{-2}$) for each run. For detail, we refer to [28]. Since W1 > W2, pedestrian movement in front of bottleneck is restrained but the movement inside is like in open corridor.

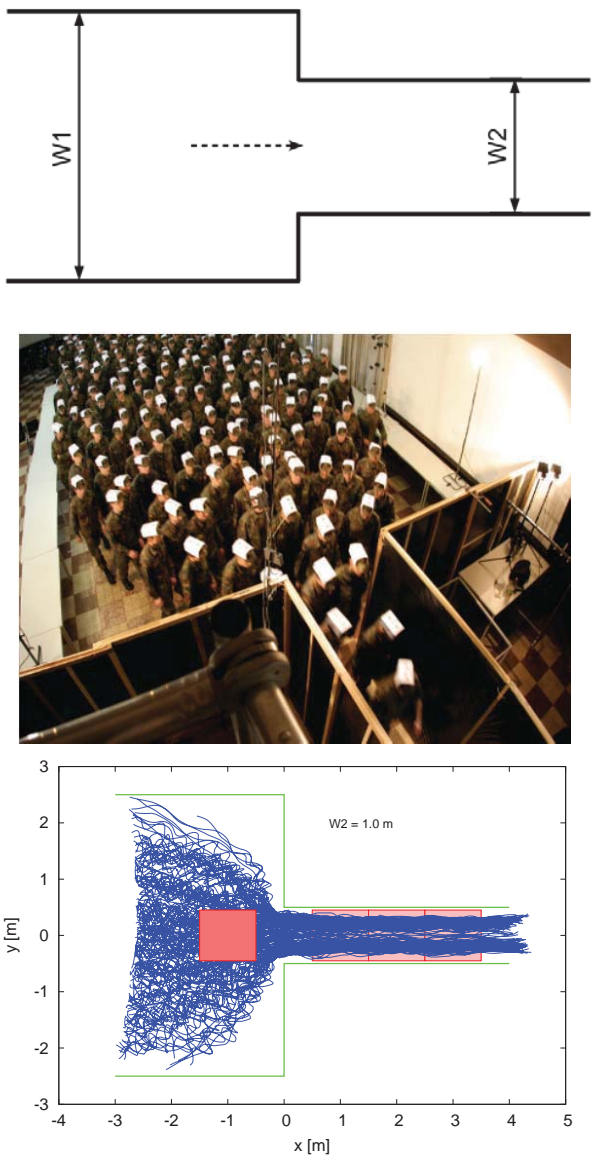

Figure 4. Sketch, snapshpot and pedestrian trajectories of the bottleneck experiment.

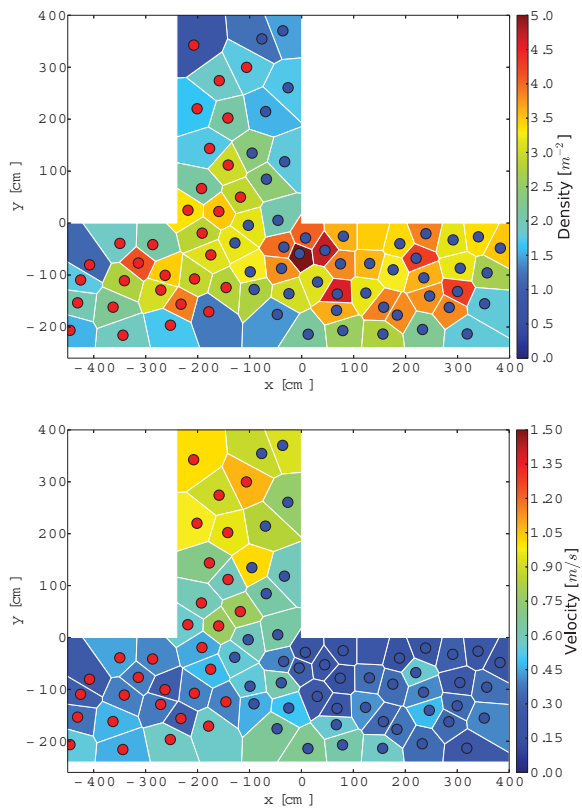

Figure 5. Density and velocity distribution over space at a fixed point in time obtained from the Voronoi method. The colors in the graph represent different level of density and velocity. The circles show the postions of pedestrian .

III. RESULTS AND ANALYSIS

In this section, the measurement method and pedestrian characteristics in the experiments are presented.

*A. Measurement method*

The Voronoi method is used to analyze the above mentioned experiments. Voronoi diagrams can be generated for every set of pedestrian positions at a fixed point in time. It contains a set of Voronoi cells which can be interpreted as the personal space belonging to each pedestrian (see Figure 5). Then the velocity and density distribution over each cell at a given time can be defined by the corresponding pedestrian's instantaneous velocity and inverse of the personal space respectively. As a result, the Voronoi density ($\rho$) and velocity (v) in a given area can be obtained based on the distributions. In our analysis, the specific flows are calculated by $J_s = \rho v$. The details for the Voronoi method can found in [27, 29].

*B. Density-flow relationships*

In the straight corridor experiment, we study the pedestrian characteristics in the measurement area as shown in Figure 1. Figure 6 shows the density-specific flow relationship in the corridor with width W1 = 1.8 m and 3.0 m.

The diagrams for the two widths agree well and the detailed comparison can be found in [27]. To studies the boundary effect we use different symbols to show the results from the experimental settings. It can be seen that the diagram can be divided into two parts. For $\rho < 2$ m$^{-2}$ the specific flow increases with the increasing density, whereas it decreases when the density is higher than 2 m$^{-2}$. To get higher densities in corridor the exit width W2 was decreased. All data for $\rho < 2$ m$^{-2}$ are obtained from experimental settings with W1 = W2, see green stars in Fig. 6. When the exit is narrower than the corridor, the outflow is restrained. The maximal density observed under this condition is about 4 m$^{-2}$, where the specific flow decreases to 0.5 (m.s)$^{-1}$, se red dots in Fig 6.

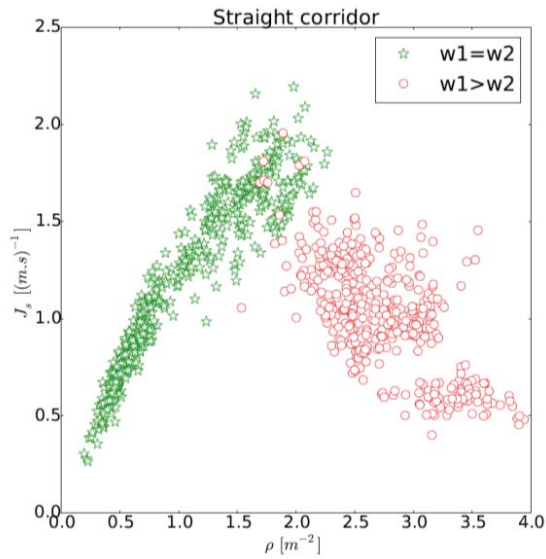

Figure 6. Density-specific flow relationships from the experiment in 1.8 m and 3.0 m straight corridor.

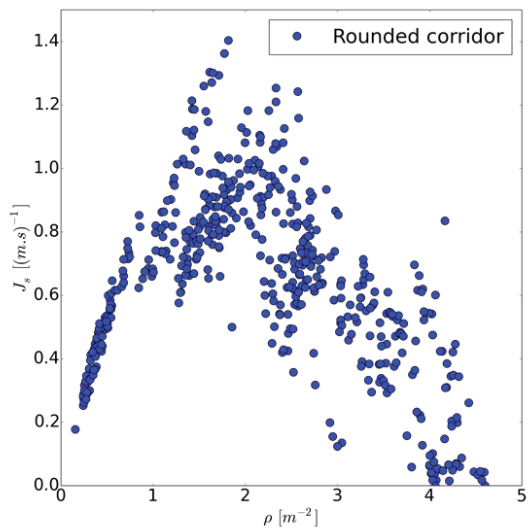

Figure 7. Density-specific flow relationships from the experiment in 1.8 m rounded corridor with periodical boundary condition.

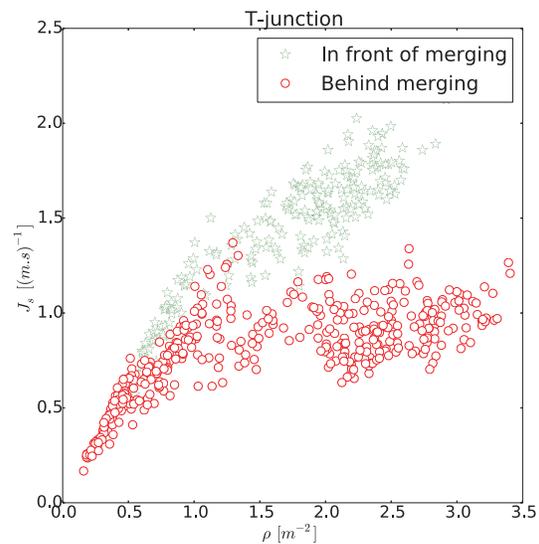

Figure 8. Density-specific flow relationships from the experiment in a 2.4 m wide T-junction.

Figure 7 shows the data from the rounded corridor (W =1.8 m) which is a system with periodic boundary conditions. Similar to the results from straight corridor, the diagram can also be divided into two parts from $\rho = 2$ m$^{-2}$. However, the maximal specific flow in this scenario is about 1.2 (m.s)$^{-1}$ which is clearly smaller than 2 (m.s)$^{-1}$ in straight corridor. Furthermore, pedestrians can hardly move when the density is around 4 m$^{-2}$ and the specific flow reaches zero in this system.

In Figure 8 we compare the density-flow relationship obtained in T-junction with corridor width W=2.4 m. In this experiment two branches merge into one stream. Since the widths of the three parts of the corridor are the same and turning behavior of pedestrian occurs during merging, the flows in the branches are restrained with the incoming streams. Due to the symmetrical setting of the geometry, the diagrams of the two branches in front of the merging match well. Moreover the specific flow shows a plateau for densities $\rho > 2.0$ m$^{-2}$. However, the diagram of the joined stream behind the merging presents obvious discrepancies. Pedestrian movement behind merging is similar to that in open corridor and the flow is independent of the outflow. The specific flow behind the merging increases continuously with the density till 2.5 m$^{-2}$ in the experiment and is significantly higher than the one measured in front of the merging of the streams at the same density.

Regarding pedestrian experiment of bottleneck a similar analysis is done. To exclude the influence of boundary effect on the measured results, $1 \times W2$ m$^2$ measurement areas are chosen. The location of these areas is shown in Figure 4. We choose one measurement area in front of the bottleneck and three inside the bottleneck. As shown in Figure 9, the densities range from 3 to 6 m$^{-2}$ in front of the bottleneck because of congestions and the specific flow also seems constant. The specific flow inside the bottleneck increases continuously to 2.5 (m.s)$^{-1}$ with increasing density till 3.5 m$^{-2}$.

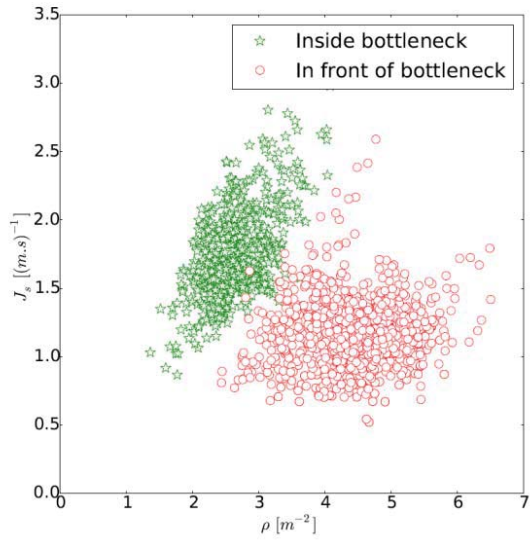

Figure 9. Density-specific flow relationships from bottleneck experiment.

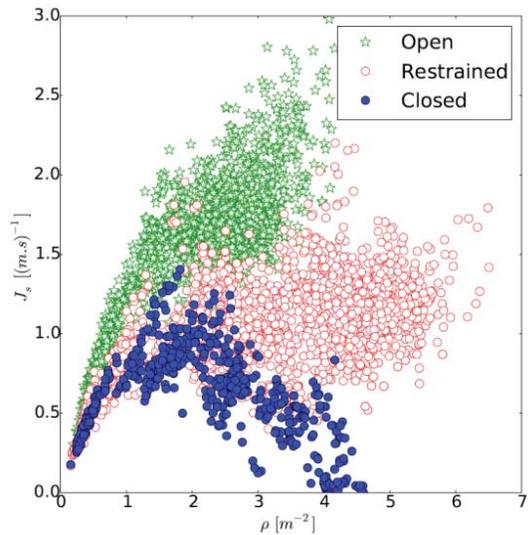

Figure 10. Comparison of density-specific flow relationships in corridor with different boundary conditions. 'Open' means that the exit of the corridor has the same width as the corridor and the outflow is not restrained. 'Restrained' means the flow in the corridor is influenced by narrowings, corners or mergings. The rounded corridor with closed and periodical boundary condition is named 'Closed'.

Based on the above analysis, it seems that the density-flow relationship strongly depends on the boundary condition of the system and no unique relation can be found for the complete system. In this point of view, we put all these diagrams together in Figure 10 to make a clear comparison. The specific flow in open corridor is always higher than other boundary conditions. When the density is smaller than 2 $m^{-2}$, the specific flows match for the closed periodical system and outflow restrained system. However, the flow keeps constant in restrained system for $2.0 < \rho < 6.0$ $m^{-2}$ which is the maximum outflow of corridor. While in closed periodical systems the specific flow decreases continuously from 1.0 $(m.s)^{-1}$ and reaches zero around $\rho = 4.0$ $m^{-2}$. However, all methods of guidelines and handbooks like SFPE [6], Predtechenskii and Milinskii [7] and Weidmann [8] et al are based on the assumption that there is a unique density-flow relation for corridors with or without bottlenecks and other narrowing. The comparison shows strong limitation of such assumption and raise questions to these methods.

## IV. SUMMARY

In this study, we present series of well-controlled laboratory experiments performed in straight and rounded corridors, T-junctions and bottlenecks. We recorded the whole processes of the experiments by two cameras and the pedestrian trajectories are extracted automatically using the software *PeTrack*. The Voronoi method is chosen to analyze the experimental data. The density-flow relationships of pedestrian streams in different scenarios are shown and discrepancies are observed. To investigate the influences of boundary conditions on the relationship, the systems are divided into three classes including open corridor, rounded corridor with periodical boundary condition as well as outflow restrained corridor. It is found that no unique relationship can be applied to the complete system and the differences mainly appear when the density is larger than 2 $m^{-2}$. With the increasing density from 2.0 to 4.0 $m^{-2}$, the specific flow in open corridor increases from 1.5 to 2.5 $(m.s)^{-1}$ but decreases from 1.0 $(m.s)^{-1}$ to zero in the system with periodic boundary conditions system. Specifically it shows a plateau at the density range when the outflow is restrained. These empirical data will be useful for the facility design and model calibrations.